\documentclass[reqno]{amsart}

\usepackage[T1]{fontenc}
\usepackage{mathptmx}
\usepackage{amssymb,amsthm}
\usepackage{graphicx}
\usepackage{url}

\theoremstyle{plain}
\newtheorem{theorem}{Theorem}

\newtheorem{lemma}[theorem]{Lemma}

\theoremstyle{definition}

\newtheorem{problem}[theorem]{Problem}
\newtheorem{assumption}[theorem]{Assumption}

\theoremstyle{remark}

\newcommand{\samp}[1]{{\mathcal S}_{#1}}
\newcommand{\hold}[1]{{\mathcal H}_{#1}}
\newcommand{\lift}{\mathcal{L}}
\newcommand{\Real}{\mathbb{R}}
\newcommand{\Rp}{\Real_{+}}
\newcommand{\aopt}{\alpha^{\mathrm{opt}}}
\newcommand{\vaopt}{\vc{\alpha}^{\mathrm{opt}}}
\newcommand{\dJ}{\nabla_{\!\vc{\alpha}}J}
\newcommand{\dnJ}{\nabla_{\!\vc{\alpha}[n]}J}

\newcommand{\A}{\Psi}

\newcommand{\C}{\mathcal{C}}
\newcommand{\D}{\mathcal{D}}

\newcommand{\Z}{\mathbb{Z}}
\newcommand{\Zp}{\mathbb{Z}_{+}}

\newcommand{\e}{\mathrm{e}}
\newcommand{\jj}{\mathrm{j}}
\newcommand{\Fh}{{\mathcal{F}}_h}

\newcommand{\vc}[1]{{\boldsymbol{#1}}}

\newcommand{\dd}{\mathrm{d}}
\newcommand{\dx}{x_{\dd}}
\newcommand{\dy}{y_{\dd}}

\begin{document}
\author[M. Nagahara]{Masaaki Nagahara}
\author[K. Hamaguchi]{Ken-ichi Hamaguchi}
\author[Y. Yamamoto]{Yutaka Yamamoto}
\address{The authors are with
Kyoto University, Graduate School of Informatics, Kyoto, Japan.
The corresponding author is M. Nagahara (nagahara@ieee.org).}
\title[ANC with Sampled-Data Filtered-$x$ Adaptive Algorithm]{\Large Active Noise Control with Sampled-Data Filtered-$x$ Adaptive Algorithm}

\maketitle

\begin{abstract}
Analysis and design of filtered-$x$ adaptive algorithms are conventionally 
done by assuming that the transfer function in the secondary path is 
a discrete-time system. However, in real systems such as active noise control, 
the secondary path is a continuous-time system. 
Therefore, such a system should be analyzed and designed as a hybrid system 
including discrete- and continuous- time systems and AD/DA devices. 
In this article, we propose a hybrid design taking account of continuous-time behavior 
of the secondary path via lifting (continuous-time polyphase decomposition) technique 
in sampled-data control theory.
\end{abstract}

\section{Introduction}
\label{sec:intro}
Recent development of digital technology enables us
to make digital signal processing (DSP) systems
much more robust, flexible, and cheaper
than analog systems.
Owing to the recent digital technology,
advanced adaptive algorithms
with fast DSP devices
are used in \emph{active noise control} (ANC) systems~\cite{EllNel93,MeuVerEll02};
air conditioning ducts~\cite{KobFuj08},
noise canceling headphones~\cite{KuoMitGan06},
and automotive applications~\cite{ShoKnu96},
to name a few.

\begin{figure}[t]
\centering
\includegraphics[width=0.8\linewidth]{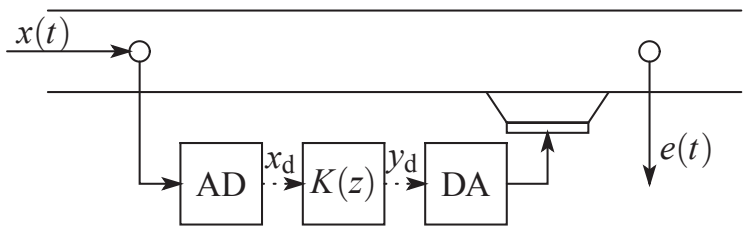}
\caption{Active noise control system}
\label{fig:anc}
\end{figure}

Fig.~\ref{fig:anc} shows a standard active noise control system.
In this system, $x(t)$ represents continuous-time noise 
which we want to eliminate during
it goes through the duct.
Precisely, we aim at diminishing the noise at the point {\tt C}.
For this purpose, we set a loudspeaker near the point {\tt C}
which emits anti-phase sound signals to cancel the noise.
Since the noise is unknown in many cases,
it is almost impossible to determine anti-phase signals
\emph{a priori}.
Hence, we set a microphone at the point {\tt A}
to measure the continuous-time noise,
and adopt a digital filter $K(z)$ with
AD (analog-to-digital) and DA (digital-to-analog) devices.
Namely, the continuous-time signal $x(t)$
is discretized to produce a discrete-time signal $\dx$,
which is processed by the digital filter $K(z)$ to produce 
another discrete-time signal $\dy$.
Then a DA converter and a loudspeaker at the point {\tt B}
are used to emit anti-phase signals
to cancel the noise in the duct.

In active noise control, it is important to compensate the distortion
by the transfer characteristic of the secondary path (from {\tt B} to {\tt C}).
To compensate this, a standard adaptive algorithm uses 
a filtered signal of the noise $x$, and is called 
\emph{filtered-x algorithm}~\cite{Mor80}.
This filter is usually chosen by a discrete-time
model of the secondary path~\cite{Mor80,EllNel93}.
Consequently, the adaptive filter $K(z)$ optimizes
the norm (or the variance in the stochastic setup) of the
discretized signal $e(nh)$,
$n=0,1,2,\ldots$ where $h$ is the sampling period of AD and DA device.
This is proper if the secondary path is also a discrete-time system.
However, in reality, the path is a \emph{continuous-time} system,
and hence the optimization should be executed taking account of 
the behavior of the continuous-time error signal $e(t)$.
Such an optimization may seem to be difficult because
the system is a \emph{hybrid} system
containing both continuous- and discrete-time signals.

Recently, several articles have been devoted to the design
considering a continuous-time behavior.
In~\cite{SonGonKuo05},
a hybrid controller containing an analog filter and a digital adaptive filter
has been proposed.
Owing to the analog filter, a robust performance is attained
against the variance of the secondary path.
However, an analog filter is often unwelcome because of its poor reliability
or maintenance cost.
Another approach has been proposed in~\cite{MeuVerEll02}.
In this paper, they assume
that the noise $x(t)$ is a linear combination of
a finite number of sinusoidal waves.
Then the adaptive algorithm is executed in the frequency domain
based on the frequency response of the continuous-time secondary path.
This method is very effective if we {\it a priori} know
the frequencies of the noise.
However, unknown signal with other frequencies cannot be eliminated.
If we prepare adaptive filters considering many frequencies
to avoid such a situation,
the complexity of the controller will be very high.

The same situation has been considered in control systems theory.
The modern \emph{sampled-data control theory}~\cite{CheFra}
has been developed in 90's~\cite{Yam94},
which gives an exact design/analysis method for hybrid systems
containing continuous-time plants and discrete-time controllers.
The key idea is \emph{lifting}.
Lifting is a transformation of continuous-time signals to an
infinite-dimensional (i.e., function-valued)  discrete-time signals.
The operation can be interpreted as a 
\emph{continuous-time polyphase decomposition}.
In multirate signal processing, the (discrete-time) polyphase decomposition 
enables the designer to perform all computations at the
lowest rate~\cite{Vai}.
In the same way, by lifting, 
continuous-time signals or systems can be represented in the discrete-time domain
with no errors.

The lifting approach is recently applied to 
digital signal processing~\cite{KasYamNag04,NagYam05,YamNagKha12},
and proved to provide an effective method for digital filter design.
Motivated these works,
this article focuses on
a new scheme of filtered-$x$ adaptive algorithm
which takes account of the continuous-time behavior.
More precisely, we define the problem of active noise control
as design of the digital filter which minimizes
a continuous-time cost function.
By using the lifting technique,
we derive the Wiener solution for this problem,
and a steepest descent algorithm based on the Wiener solution.
Then we propose an LMS (least mean square) type algorithm
to obtain a causal system.
The LMS algorithm involves an integral computation on a finite interval,
and we adopt an approximation based on lifting representation.
The approximated algorithm can be easily executed 
by a (linear, time-invariant, and finite dimensional) digital filter.

The paper is organized as follows:
Section \ref{sec:FXLMS} formulates the problem of active noise control.
Section \ref{sec:SD} gives the Wiener solution,
the steepest descent algorithm, and the LMS-type algorithm
with convergence theorems.
Section \ref{sec:approximation} proposes an approximation method
for computing an integral of signals for the LMS-type algorithm.
Section \ref{sec:simulation}
shows simulation results to illustrate
the effectiveness of the proposed method.
Section \ref{sec:conc} concludes the paper.

\subsection*{Notation}
\begin{description}
\item[$\Real$, $\Rp$]: the sets of real numbers and non-negative real numbers, respectively.
\item[$\Z$, $\Zp$]: the sets of integers and non-negative integers, respectively.
\item[$\Real^n$, $\Real^{n\times m}$]: the sets of
$n$-dimensional vectors and $n\times m$ matrices over $\Real$, respectively.
\item[$L^2$, $L^2[0,h)$]: the sets of all square integrable functions on $\Rp$ and $[0,h)$, respectively.
\item[$M^\top$]: transpose of a matrix $M$.
\item[$\overline{a}$]: the complex conjugate of a complex number $a$
\item[$s$]: the symbol for Laplace transform
\item[$z$]: the symbol for $Z$ transform
\end{description}

\section{Problem Formulation}
\label{sec:FXLMS}
\begin{figure}[t]
\centering
\includegraphics[width=0.8\linewidth]{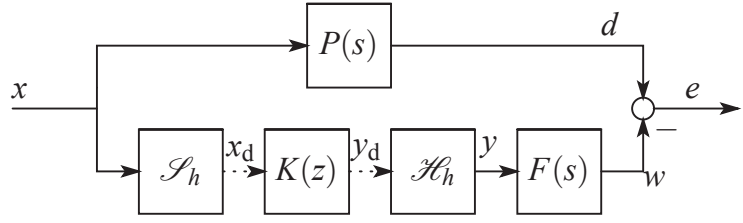}
\caption{Block diagram of active noise control system}
\label{fig:anc_block}
\end{figure}

In this section, we formulate the design problem
of active noise control.
Let us consider the block diagram shown in Fig.~\ref{fig:anc_block}
which is a model of the active noise control system  shown in Fig.~\ref{fig:anc}.
In this diagram,
$P(s)$ is the transfer function of the primary path
from {\tt A} to {\tt C} in Fig.~\ref{fig:anc}.
The transfer function of the secondary path from {\tt B} to {\tt C} is represented by $F(s)$.
Note that $P(s)$ and $F(s)$ are continuous-time systems.
We model the AD device by the ideal sampler $\samp{h}$ with a sampling period $h$
defined by
\[
  (\samp{h}x)[n]:=x(nh),
       \quad n\in\Zp.
\]
That is, the ideal sampler $\samp{h}$ converts continuous-time signals to discrete-time signals.
Then, the DA device is modeled by the zero-order hold $\hold{h}$ with the same period $h$
defined by
\[
           (\hold{h}y)(t):=\sum_{n=0}^\infty \phi_0(t-nh)y[n], \quad t\in [0,\infty),
\]
where $\phi_0(t)$ 
is the zero-order hold function or the box function defined by
\[
\phi_0(t):= \begin{cases}1, \quad 
         t \in [0,h),\\ 0, \quad \text{otherwise}.\end{cases}
\]
That is, the zero-order hold $\hold{h}$ converts
discrete-time signals to continuous-time signals.

With the setup,
we formulate the design problem as follows:
\begin{problem}
Find the optimal FIR (finite impulse response) filter
\[
K(z) = \sum_{k=0}^{N-1} \alpha_k z^{-k}
\]
which minimizes the continuous-time cost function
\begin{equation}
\label{eq:J}
J=\int_0^\infty e(t)^2 \dd t.
\end{equation}
\end{problem}
Instead of the conventional adaptive filter design~\cite{Hay},
this problem deals with the continuous-time behavior of the error signal $e(t)$.
To solve such a hybrid problem 
(i.e., a problem for a mixed continuous- and discrete-time system), 
we introduce the lifting approach based on the sampled-data control theory~\cite{CheFra}.

In what follows, we assume the following:
\begin{assumption}
The following properties hold:
\begin{enumerate}
\item The noise $x$ is unknown but causal, that is,
$x(t)=0$ if $t<0$, and belongs to $L^2$.
\item The primary path $P(s)$ is unknown, but proper and stable.
\item The secondary path $F(s)$ is known, proper and stable.
\end{enumerate}
\end{assumption}

\section{Sampled-Data Filtered-$x$ Algorithm}
\label{sec:SD}
In this section, we discretize the continuous-time 
cost function \eqref{eq:J} without any approximation,
and derive optimal filters.
We also give convergence theorems for the proposed adaptive filters.
The key idea to derive the results in this section is the 
\emph{lifting} technique~\cite{Yam94,CheFra}.

\subsection{Wiener Solution}
In this subsection, we derive the optimal filter coefficients
$\alpha_0,\alpha_1,\ldots,\alpha_{N-1}$
which minimize the cost function $J$ in \eqref{eq:J}.

First, we split the time domain $[0,\infty)$ into the union of sampling intervals
$[nh, (n+1)h)$, $n\in\Zp$, as
\[
 [0,\infty) = [0,h) \cup [h,2h) \cup [2h,3h) \cup \cdots.
\]
By this, the cost function \eqref{eq:J} is
transformed into the sum of the $L^2[0,h)$-norm of $e(t)$
on the intervals:
\begin{equation}
  J = \int_0^\infty e(t)^2 \dd t
    = \sum_{n=0}^\infty \int_{0}^{h} e(nh+\theta)^2 \dd \theta
    = \sum_{n=0}^\infty \int_{0}^{h} \vc{e}_n(\theta)^2 \dd \theta,
 \label{eq:J2}
\end{equation}
where
$\vc{e}_n(\theta) = e(nh+\theta)$, $\theta \in [0,h)$, $n\in\Zp$.
The sequence $\{\vc{e}_n\}$ of functions $\vc{e}_1,\vc{e}_2,\ldots$ on $[0,h)$
is called the \emph{lifted signal}~\cite{Yam94,CheFra} of
the continuous-time signal $e\in L^2$,
and we denote the \emph{lifting operator} by $\lift$,
that is, $\{\vc{e}_n\}=\lift e$.
In what follows, we use the notion of lifting to derive
the optimal coefficients.

Next, we assume that a state space realization is given for $F(s)$ as
\[
 F : 
  \left\{
  \begin{aligned}
  \dot{\zeta}(t) &= A\zeta(t) + By(t),\\
  w(t) &= C\zeta(t),\quad t\in\Real_+
  \end{aligned}
  \right.
\]
where $\zeta(0)=0$, $A\in\Real^{\nu\times \nu}$, $B\in\Real^{\nu\times 1}$,
and $C\in\Real^{1\times \nu}$.
By Fig.~\ref{fig:anc_block}, the continuous-time signal 
$w$ is given by 
\[
w=Fy=F\hold{h}\dy
\]
where $\dy$ is a discrete-time signal $\dy=\{\dy[n]\}$
which is produced by the filter $K(z)$.
Let $\vc{w}_n(\theta) := w(nh+\theta)$, $\theta\in[0,h)$, $n\in\Zp$
(i.e., $\{\vc{w}_n\}:=\lift w$).
Then, the sequence of functions $\{\vc{w}_n\}$ is obtained
as 
\[
 \{\vc{w}_n\} = \lift F\hold{h} \dy.
\]
Let $\Fh:=\lift F\hold{h}$.
Then the system $\Fh$ is a discrete-time system
as shown in the following lemma~\cite[Sec.~10.2]{CheFra}:
\begin{lemma}
\label{lem:Fh}
$\Fh$ is a linear time-invariant
discrete-time (infinite-dimensional) system
with the following state-space representation:
\begin{equation}
 \Fh : 
  \left\{
  \begin{aligned}
   \xi[n+1] &= A_h \xi[n] + B_h \dy[n],\\ 
   \vc{w}_n &= \C_h\xi[n] + \D_h \dy[n],\quad n\in\Zp,
  \end{aligned}
  \right.
 \label{eq:Fh}
\end{equation}
where
\begin{equation}
 \begin{split}
 A_h &:= \e^{Ah} \in \Real^{\nu\times\nu},\quad
 B_h := \int_0^h \e^{A\theta}B\dd\theta \in \Real^{\nu\times 1},\\
 \C_h &: \Real^\nu \ni \xi \mapsto C\e^{A\bullet}\xi\in L^2[0,h),\quad
 \D_h : \Real \ni \dy \mapsto \int_0^\bullet C\e^{A\tau}B\dd\tau\cdot\dy \in L^2[0,h)\\
 \end{split}
 \label{eq:FhABCD}
\end{equation}
\end{lemma}

The LTI property of $\Fh$ in Lemma \ref{lem:Fh} gives
\begin{equation}
 \begin{split}
 \{\vc{w}_n\}
  &= \Fh\{\dy[n]\}\\
  &= \Fh\left(\left\{\sum_{k=0}^{N-1}\alpha_kz^{-k}\dx[n]\right\}\right)\\
  &= \sum_{k=0}^{N-1}\alpha_k \Fh\left(\left\{z^{-k}\dx[n]\right\}\right)\\
  &= \left\{\sum_{k=0}^{N-1}\alpha_k\vc{u}_{n-k}\right\},
 \end{split}
 \label{eq:wn}
\end{equation}
where $\{\vc{u}_n\} := \Fh \{\dx[n]\}$.
Note that $\{\vc{u}_n\}$ is the lifted signal of 
the continuous-time signal $u=F\hold{h}\dx$,
that is,
\[
 \{\vc{u}_n\} = \lift(F\hold{h}\dx)=\lift u.
\]
The relation \eqref{eq:wn} gives the continuous-time
relation as
\[
 w(t) = \sum_{k=0}^{N-1} \alpha_k u(t-kh),\quad t\in\Rp.
\]
By using this relation,
we obtain the following theorem for the optimal filter.
\begin{theorem}[Wiener solution]
\label{thm:wiener}
Let $u := (F\hold{h})\dx$.
Define a matrix $\Phi$ and a vector $\vc{\beta}$ as
\[
  \Phi:=[\Phi_{kl}]_{k,l=0,1,\ldots,N-1}\in\Real^{N\times N},\quad
  \vc{\beta}:=[\beta_k]_{k=0,1,\ldots,N-1}\in\Real^{N},
\]
where for $k,l=0,1,\ldots,N-1$,
\[
  \Phi_{kl} := \int_0^\infty u(t-kh)u(t-lh)\dd t,\quad
  \beta_k := \int_0^\infty d(t)u(t-kh)\dd t.
\]
Assume the matrix $\Phi$ is nonsingular.
Then the gradient of $J$ defined in \eqref{eq:J}  is given by
\begin{equation}
 \dJ = 2\left(\Phi\vc{\alpha}-\vc{\beta}\right),\quad
 \vc{\alpha} := [\alpha_0,\alpha_1,\ldots,\alpha_{N-1}]^\top,
 \label{eq:dJ}
\end{equation}
and the optimal FIR parameter $\vaopt=[\aopt_0,\aopt_1,\ldots,\aopt_{N-1}]^\top$
which minimizes $J$ is given by
\begin{equation}
 \vaopt = \Phi^{-1}\vc{\beta}.
 \label{eq:wiener}
\end{equation}
\end{theorem} 
{\bf Proof:} Let 
$\{\vc{d}_n\}:=\lift d$. By the equations \eqref{eq:J2}, \eqref{eq:wn}, 
and $\vc{e}_n=\vc{d}_n-\vc{w}_n$, we have
\begin{multline}
  J = \sum_{n=0}^\infty \int_0^h \vc{d}_n(\theta)^2\dd \theta
    -2 \sum_{k=0}^{N-1}\alpha_k \sum_{n=0}^\infty \int_0^h \vc{d}_n(\theta)\vc{u}_{n-k}(\theta)\dd \theta\\
    + \sum_{k=0}^{N-1}\sum_{l=0}^{N-1}\alpha_k\alpha_l \sum_{n=0}^\infty \int_0^h \vc{u}_{n-k}(\theta)\vc{u}_{n-l}(\theta)\dd \theta.
\end{multline}
Computing the gradient $\dJ$ and applying the inverse lifting, 
we obtain \eqref{eq:dJ}.
Then, if the matrix $\Phi$ is nonsingular, the optimal parameter \eqref{eq:wiener} is given
by solving the Wiener-Hopf equation $\Phi\vc{\alpha}-\vc{\beta}=\vc{0}$.
\hfill $\square$

We call the optimal parameter $\vaopt$ the \emph{Wiener solution}.

\subsection{Steepest Descent Algorithm}
In this subsection, we derive the 
\emph{steepest descent algorithm} 
(SD algorithm)~\cite{Hay}
for the Wiener solution obtained in Theorem \ref{thm:wiener}.
This algorithm is a base for adaptation of the ANC system discussed in
the next subsection.

According to the identity \eqref{eq:dJ} in Theorem \ref{thm:wiener}
for the gradient of $J$,
the steepest descent algorithm is described by
\begin{equation}
 \begin{split}
  \vc{\alpha}[n+1]
   &= \vc{\alpha}[n] - \frac{\mu}{2}\dnJ\\
   &= \vc{\alpha}[n] + \mu \left(\vc{\beta} - \Phi\vc{\alpha}[n]\right),\quad n\in\Zp,
 \end{split}
 \label{eq:SDA}
\end{equation}
where $\mu>0$ is the \emph{step-size parameter}.

We then analyze the stability of the above recursive algorithm.
Before deriving the stability condition, 
we give an upper bound of the eigenvalues of the matrix $\Phi$.
\begin{lemma}
\label{lem:S}
Let $\lambda_1,\ldots,\lambda_N$ be
the eigenvalues of the matrix $\Phi$.
Let $\hat{u}$ denote the Fourier transform of $u=F\hold{h}\dx$,
and define
\[
  S(\jj\omega) 
  := \frac{1}{h} 
  \sum_{n=-\infty}^\infty 
   \left|\hat{u}\left(\jj\omega+\frac{2n\pi\jj}{h}\right)\right|^2.
\]
Then we have
\begin{equation}
0\leq\lambda_i\leq \|S\|_\infty = \sup\left\{S(\jj\omega)\mid \omega\in\left(-\tfrac{\pi}{h},\tfrac{\pi}{h}\right)\right\},
\label{eq:lambda}
\end{equation}
for $i=1,2,\ldots,N$.
\end{lemma}
{\bf Proof:} See \ref{app:lemma1}.
\hfill $\square$

By this lemma, we derive a sufficient condition on the step size $\mu$ for convergence.
\begin{theorem}[Stability of SD algorithm]
\label{thm:stability1}
Suppose that $\Phi >0$ and
the step size $\mu$ satisfies
\begin{equation}
 0 < \mu <2\|S\|_\infty^{-1}.
 \label{eq:stability1}
\end{equation}
Then the sequence $\{\vc{\alpha}[n]\}$ 
produced by the iteration \eqref{eq:SDA}
converges to the Wiener solution $\vaopt$
for any initial vector $\vc{\alpha}[0]\in\Real^N$.
\end{theorem}
{\bf Proof:}
The iteration \eqref{eq:SDA} is rewritten as
\[
 \vc{\alpha}[n+1] = (I-\mu\Phi)\vc{\alpha}[n]+\mu\vc{\beta}.
\]
Suppose $\Phi>0$.
Let $\lambda_{\max}$ denote the maximum eigenvalue of $\Phi$.
Then  $\lambda_{\max}>0$ since $\Phi>0$.
The condition \eqref{eq:stability1} and the inequality 
\eqref{eq:lambda} in Lemma \ref{lem:S} give
$0<\mu<2\lambda_{\max}^{-1}$,
which is equivalent to $\left|1-\mu\lambda_i\right|<1$, $i=1,2,\ldots,N$.
It follows that the eigenvalues of the matrix $I-\mu\Phi$ lie in the
open unit disk in the complex plane,
and hence the iteration \eqref{eq:SDA} is asymptotically stable.
The final value 
\[
 \vc{\alpha}_\infty := \lim_{n\rightarrow\infty}\vc{\alpha}[n]
\]
of the iteration is clearly given by the solution
of the equation $\Phi\vc{\alpha}_\infty=\vc{\beta}$.
Thus, since $\Phi>0$,  we have $\vc{\alpha}_\infty=\Phi^{-1}\vc{\beta}=\vaopt$.
\hfill $\square$

\subsection{LMS-type Algorithm}
The steepest descent algorithm assumes that
the matrix $\Phi$ and the vector $\vc{\beta}$ are known {\it a priori}.
That is, the noise $\{x(t)\}_{t\in\Rp}$ and the primary path 
$P(s)$ are assumed to be known.
However, in practice, 
the noise $\{x(t)\}_{t\in\Rp}$ cannot be fixed
before we run the ANC system.
In other words, the ANC system should be \emph{noncausal}
for running the steepest descent algorithm.
Moreover, we cannot produce arbitrarily noise $\{x(t)\}_{t\in\Rp}$
(this is why $x$ is \emph{noise}),
we cannot identify the primary path $P(s)$.
Hence, the assumption is difficult to be satisfied.

In the sequel, we can only use data
up to the present time for causality
and we cannot use the model of $P(s)$.
Under this limitation, we
propose to use an LMS-type adaptive algorithm
using the filtered noise $u=F\hold{h}\dx$ and 
the error $e$ up to the present time.

\begin{figure}[t]
\centering
\includegraphics[width=0.8\linewidth]{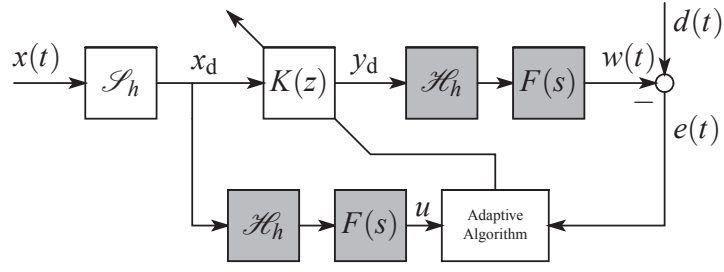}
\caption{Sampled-data filtered-$x$ adaptive algorithm}
\label{fig:fx}
\end{figure}

First, by the equation \eqref{eq:wn} and the relation $e=d-w$,
we have
\[
 \frac{\partial J}{\partial \alpha_k}
  = -2 \biggl(\beta_k-\sum_{l=0}^{N-1}\Phi_{kl}\alpha_l\biggr)
  = -2 \int_0^\infty e(t) u(t-kh)\dd t,\quad
  k=0,1,\ldots,N-1.
\]
Based on this, we propose the following adaptive algorithm:
\begin{equation}
 \vc{\alpha}[n+1] = \vc{\alpha}[n] + \mu \vc{\delta}[n],\quad n\in\Zp,
 \label{eq:LMS}
\end{equation}
where $\vc{\delta}[n]=\bigl[\delta_0[n],\delta_1[n],\ldots,\delta_{N-1}[n]\bigr]^\top$
with
\[
  \delta_k[n] := \int_0^{nh} e(t)u(t-kh)\dd t,\quad
  k=0,1,\ldots,N-1.
\]
The update direction vector $\vc{\delta}[n]$ can be recursively computed by
\begin{equation}
 \vc{\delta}[n+1] = \vc{\delta}[n] + \int_{nh}^{(n+1)h} e(t)\vc{u}(t) \dd t,\quad n\in\Zp,
 \label{eq:dn}
\end{equation}
where
\[
 \vc{u}(t) := \biggl[u(t), u(t-h),\ldots, u\bigl(t-(N-1)h\bigr)\biggr]^\top.
\]
This means that to obtain the vector $\vc{\delta}[n]$ one needs
to measure the error $e$ and the signal $u=F\hold{h}\dx$ on the interval $[(n-1)h,nh)$
and compute the integral in \eqref{eq:dn}.
We call this scheme the
\emph{sampled-data filtered-$x$ adaptive algorithm}.
The term ``sampled-data'' comes from the use of
sampled-data $\dx$ of the continuous-time signal $x$.
The sampled-data filtered-$x$ adaptive algorithm
is illustrated in Fig.~\ref{fig:fx}.
As shown in this figure,
in order to run the adaptive algorithm,
we should use the signal $u$ which is ``filtered'' $\dx$ by $F\hold{h}$,
and also use the error signal $e$.

To analyze the convergence of the iteration,
we consider the following autonomous system:
\begin{equation}
 \vc{\alpha}[n+1] = \bigl(I-\mu\Phi[n]\bigr)\vc{\alpha}[n],\quad n\in\Zp,
 \label{eq:auto}
\end{equation}
where $\Phi[n]=\bigl[\Phi_{kl}[n]\bigr]_{k,l=0,1,\ldots,N-1}$ with
\[
 \Phi_{kl}[n] := \int_0^{nh} u(t-kh)u(t-lh)\dd t.
\]
Then we have the following lemma:
\begin{lemma}
\label{lem:stability2}
Suppose the following conditions:
\begin{enumerate}
\item The sequence $\{\Phi[n]\}$ is uniformly bounded, that is,
there exists $\gamma>0$ such that
\[
 \|\Phi[n]\| \leq \gamma,\quad \forall n\in\Zp.
\]
\item The step-size parameter $\mu$ satisfies
\[
 0<\mu<2\biggl(\max_{n\in\Zp}\lambda_{\max}\bigl(\Phi[n]\bigr)\biggr)^{-1},
\]
where $\lambda_{\max}\bigl(\Phi[n]\bigr)$ is the maximum eigenvalue of $\Phi[n]$.
\item The sequence $\{\mu\Phi[n]\}$ is slowly-varying, that is,
there exists a sufficiently small $\epsilon>0$ such that
\[
 \bigl\|\mu\bigl(\Phi[n]-\Phi[n-1]\bigr)\bigr\|\leq\epsilon,\quad \forall n\in\Zp.
\]
\end{enumerate}
Then the autonomous system \eqref{eq:auto} is uniformly exponentially stable%
\footnote{
The system \eqref{eq:auto} is said to be 
\emph{uniformly exponentially stable}~\cite{Rug} if
there exist a finite positive constant $c$ and a constant $0\leq r<1$
such that for any $n_0$ and $\vc{\alpha}_0=\vc{\alpha}[0]\in\Real^N$,
the corresponding solution satisfies
$\|\vc{\alpha[n]}\|\leq cr^{n-n_0}\|\vc{\alpha}_0\|$
for all $n\geq n_0$.}.
\end{lemma}
{\bf Proof:} See \ref{app:lemma2}.
\hfill $\square$

By Lemma \ref{lem:stability2}, we have the following theorem:
\begin{theorem}[Stability of LMS algorithm]
\label{thm:stability2}
Suppose the conditions 1--3 in Lemma \ref{lem:stability2}.
Then the sequence $\{\vc{\alpha}[n]\}$ converges
to the Wiener solution $\vaopt$.
\end{theorem}
{\bf Proof:}
Let $\vc{\beta}[n]:=\bigl[\beta_k[n]\bigr]_{k=0,1,\ldots,N-1}\in\Real^{N}$ with
\[
 \beta_k[n] := \int_0^{nh} d(t)u(t-kh)\dd t.
\]
Put
$\vc{c}[n]:=\vc{\alpha}[n]-\vaopt$
and $\vc{q}[n]:= \vc{\beta}[n]-\Phi[n]\vaopt$.
Then, $\Phi[n]\rightarrow\Phi$ and $\vc{\beta}[n]\rightarrow\vc{\beta}$ as
$n\rightarrow\infty$,
and hence 
\begin{equation}
 \vc{q}[n]\rightarrow\infty~\text{ as }~n\rightarrow\infty.
 \label{eq:qn}
\end{equation}
By Lemma \ref{lem:stability2}, the autonomous system \eqref{eq:auto}
is uniformly exponentially stable and from \eqref{eq:qn} it follows that
$\vc{c}[n]\rightarrow\vc{0}$ as $n\rightarrow\infty$.
Thus, we have
$\vc{\alpha}[n]\rightarrow\vaopt$ as $n\rightarrow\infty$.
\hfill $\square$

\section{Approximation Method}
\label{sec:approximation}
To run the algorithm \eqref{eq:LMS} with \eqref{eq:dn},
we have to calculate the
integral in \eqref{eq:dn}.
It is usual that the error signal $e$ is given as
sampled data, and hence
the exact value of this integral
is difficult to obtain in practice.
Therefore, we introduce an approximation method 
for this computation.

First, we split the interval $[0,h)$ into $L$ short intervals as
\[
[0,h) = [0,h/L)\cup[h/L,2h/L)\cup\cdots\cup[h-h/L, h).
\]
Assume that the error $e$ is constant on each short interval.
Then we have,
\[
  \int_{nh}^{(n+1)h} e(t)u(t-kh)\dd t
  = \sum_{l=0}^{L-1} \int_{lh/L+nh}^{(l+1)h/L+nh} e(t)u(t-kh)\dd t
  = \vc{e}[n]^\top \vc{U}[n-k],
\]
where
\[
  \vc{e}[n] := \begin{bmatrix}e(nh)\\e(h/L+nh)\\\vdots\\e(h-h/L+nh)\end{bmatrix},\quad
  \vc{U}[n] := \begin{bmatrix}\int_0^{h/L}u(\theta+nh)\dd\theta\\\int_{h/L}^{2h/L}u(\theta+nh)\dd\theta\\\vdots\\\int_{(L-1)h/L}^{h}u(\theta+nh)\dd\theta\end{bmatrix}.
\]
Then the integral in  $\vc{U}[n]$ can be computed via
the state-space representation of $\Fh$ given in \eqref{eq:Fh}.
In fact, $\vc{U}[n]$ can be computed by the following digital filter:
\[
F_h \left\{
\begin{aligned}
\eta[n+1] &= A_h \eta [n] + B_h \dx[n],\\
\vc{U}[n] &= C_h \eta[n] + D_h \dx[n],\quad n\in\Z_+
\end{aligned}
\right.
\]
where $A_h$ and $B_h$ are given in \eqref{eq:FhABCD}, 
$C_h$ and $D_h$ are matrices defined by
\[
C_h :=	\begin{bmatrix}\int_{0}^{h/L}Ce^{A\theta}\dd\theta\\\int_{h/L}^{2h/L}Ce^{A\theta}\dd\theta\\\vdots\\\int_{(L-1)h/L}^{h}Ce^{A\theta}\dd\theta\end{bmatrix},\quad
D_h :=	\begin{bmatrix}\int_{0}^{h/L}\int_0^\theta Ce^{A\tau}\dd\tau\dd\theta\\\int_{h/L}^{2h/L}\int_0^\theta Ce^{A\tau}\dd\tau\dd\theta\\\vdots\\\int_{(L-1)h/L}^{h}\int_0^\theta Ce^{A\tau}\dd\tau\dd\theta\end{bmatrix}.
\]
Note that the integrals in $B_h$, $C_h$, and $D_h$ can be effectively
computed by using matrix exponentials~\cite{Loa78,CheFra}.

\begin{figure}[t]
\centering
\includegraphics[width=0.8\linewidth]{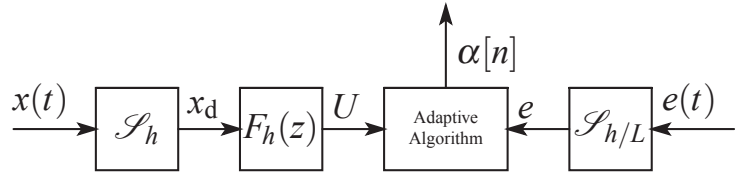}
\caption{filtered-$x$ adaptive scheme}
\label{fig:lms_alg}
\end{figure}

Let us summarize  the proposed adaptive algorithm.
The continuous-time error $e(t)$ is sampled with the fast sampling period $h/L$
and blocked to become the discrete-time signal $\vc{e}[n]$,
and the signal $x(t)$ is sampled with the sampling period $h$
to become $\dx[n]$.
Then the sampled signal $\dx$ is filtered by $F_h(z)$
and the signal $\vc{U}[n]$ is obtained.
By using $\vc{e}[n]$ and $\{\vc{U}[n],\vc{U}[n-1],\ldots,\vc{U}[n-N+1]\}$,
we update the filter coefficient $\vc{\alpha}[n]$ by 
\eqref{eq:LMS} and \eqref{eq:dn}
with
\[
 \int_{nh}^{(n+1)h} e(t)\vc{u}(t)\dd t \approx \begin{bmatrix}\vc{e}[n]^\top\vc{U}[n]\\\vc{e}[n]^\top\vc{U}[n-1]\\\vdots\\\vc{e}[n]^\top\vc{U}[n-N+1]\end{bmatrix}.
\]
We show the proposed  adaptive scheme in Fig.~\ref{fig:lms_alg}.

\section{Simulation}
\label{sec:simulation}
\begin{figure}[t]
\centering
\includegraphics[width=0.7\linewidth]{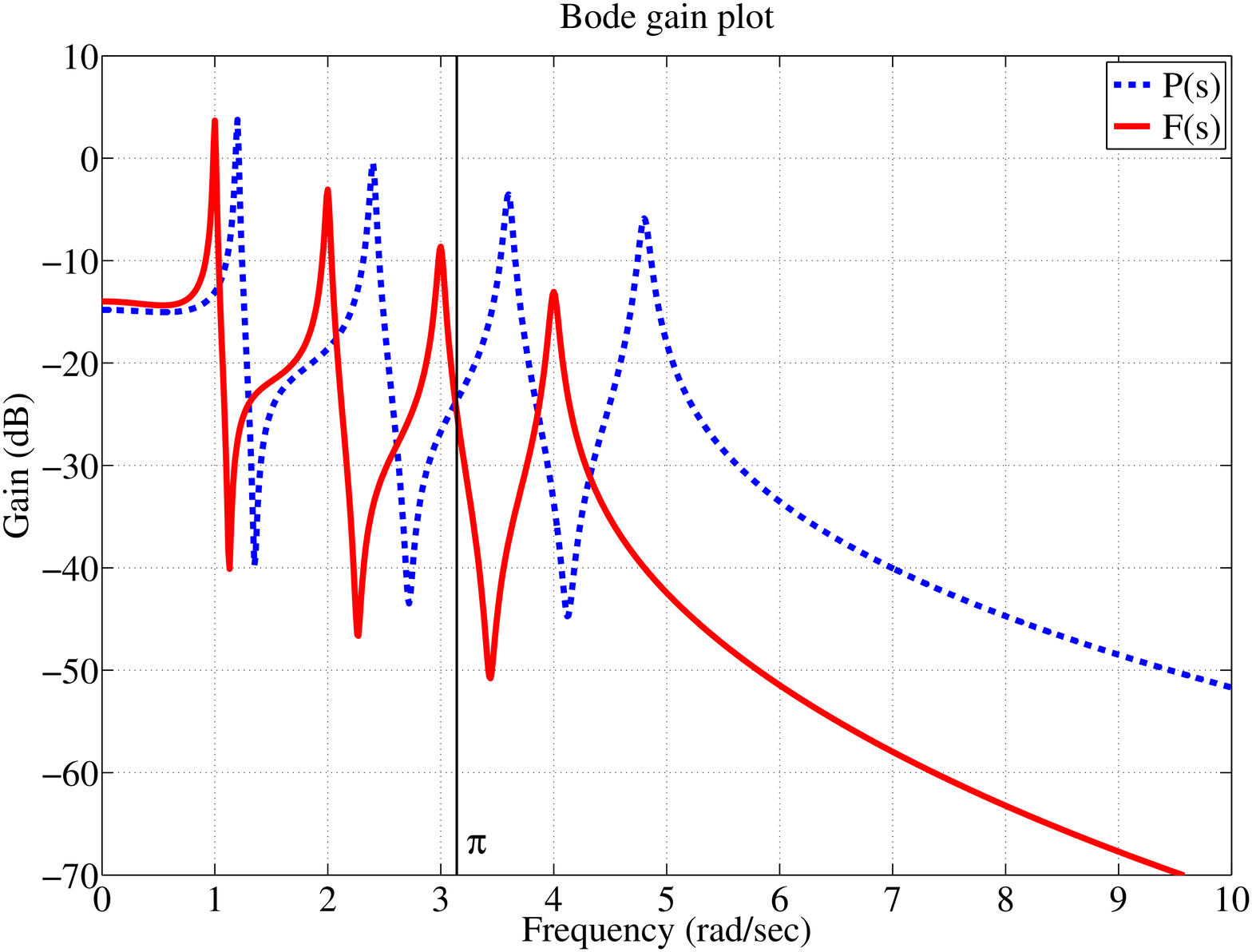}
\caption{Freqnecy response of $F(s)$ (dash) and $P(s)$ (solid). 
The vertical line indicates the Nyquist frequency $\pi$ (rad/sec).}
\label{fig:bodegain}
\end{figure}

In this section, we show simulation results
of active noise control.
The analog systems $F(s)$ and $P(s)$ are given by
\[
 \begin{split}
  F(s) &= \frac{1}{s+1.1}\cdot\frac{1}{20}\sum_{k=1}^4\frac{k^2}{s^2+2\zeta ks+k^2},\\
  P(s) &= \frac{1.2\times 1.3}{(s+1.2)(s+1.3)}\cdot\frac{1}{20}\sum_{k=1}^4\frac{(1.2k)^2}{s^2+2\zeta (1.2k)s+(1.2k)^2}.
 \end{split}
\]
The Bode gain plots of these systems are shown in 
Fig.~\ref{fig:bodegain}.
The gain $|F(\jj\omega)|$ has peaks at $\omega=1,2,3,4$ (rad/sec) and 
$|P(\jj\omega)|$ has peaks at $\omega = 1.2, 2.4, 3.6, 4.8$ (rad/sec).
We set the sampling period $h=1$ (sec) and
the fast-sampling ratio $L=8$.
Note that the systems $F(s)$ and $P(s)$ are stable and have peaks beyond the Nyquist frequency $\omega = \pi$
(rad/sec).

\begin{figure}[t]
\centering
\includegraphics[width=0.7\linewidth]{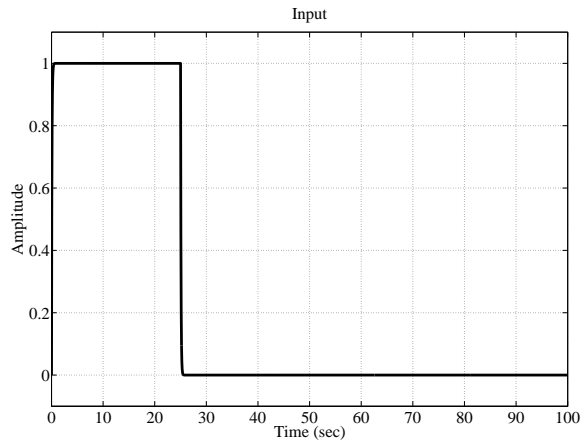}
\caption{Input signal $x(t)$ with $0\leq t\leq 100$ (sec).}
\label{fig:input}
\end{figure}

Then we run a simulation of active noise control by the proposed
method with the input signal
$x(t)$ shown in Fig.~\ref{fig:input}.
Note that the input $x(t)$ belongs $L^2$
and satisfies our assumption.
To compare with the proposed method, we also
run a simulation by a standard discrete-time LMS algorithm~\cite{EllNel93},
which is obtained by setting the fast-sampling parameter $L$ to be 1.
The step-size parameter $\mu$ in the coefficient update in
\eqref{eq:LMS} is set to be 0.1.
\begin{figure}[t]
\centering
\includegraphics[width=0.7\linewidth]{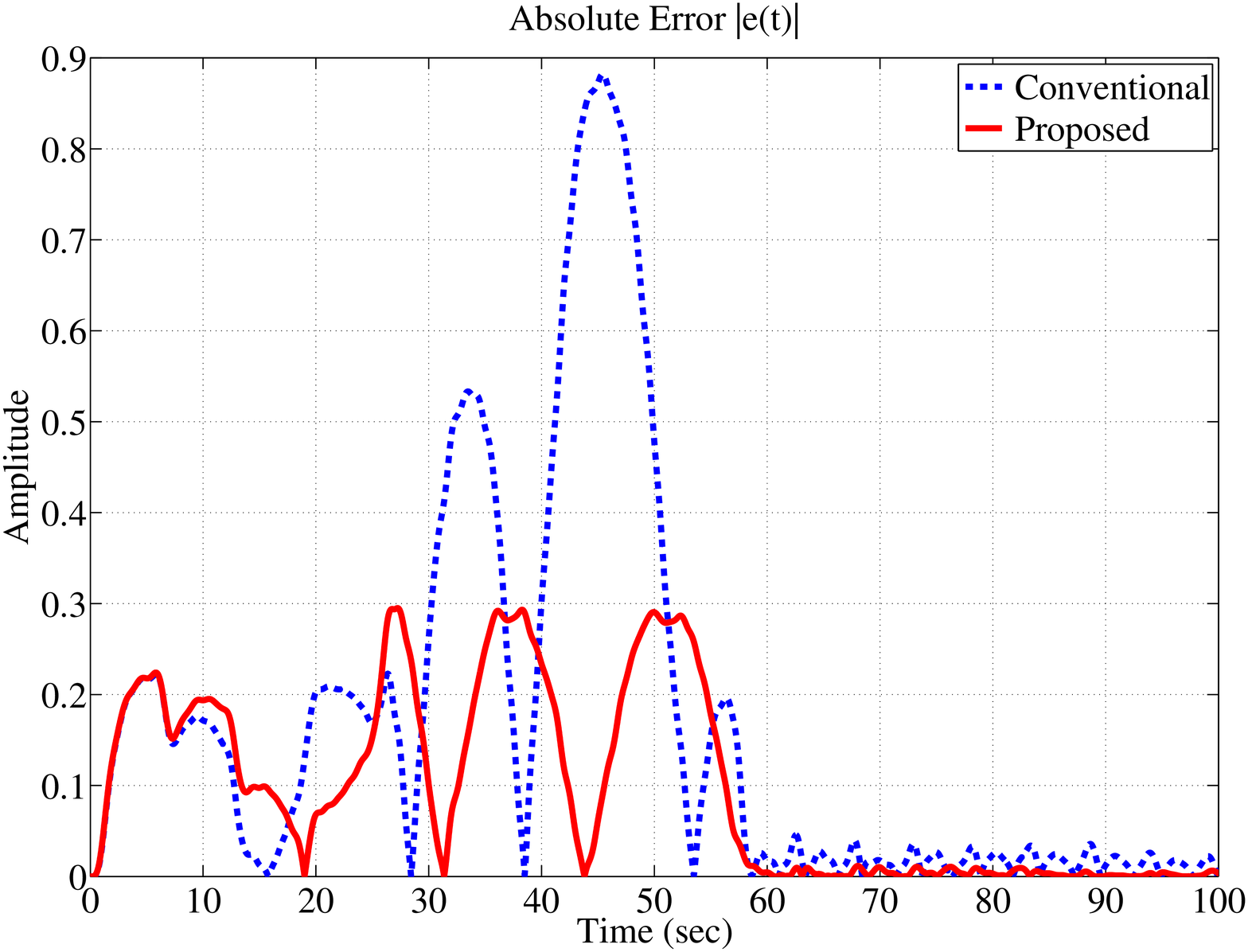}
\caption{Absolute values of error signal $e(t)$: conventional (dash) proposed (solid).}
\label{fig:errors}
\end{figure}%
Fig.~\ref{fig:errors} shows the absolute values of
error signal $e(t)$ (see Fig.~\ref{fig:anc} or Fig.~\ref{fig:anc_block}).
The errors by the conventional design is much larger than that
by the proposed method. In fact, the $L^2$ norm of the error signal $e(t)$,
$0\leq t\leq 100$ (sec) is
2.805 for the conventional method and
1.392 for the proposed one,
which is improved by about 49.6\%.
The result shows the effectiveness of our method.

\begin{figure}[t]
\centering
\includegraphics[width=0.8\linewidth]{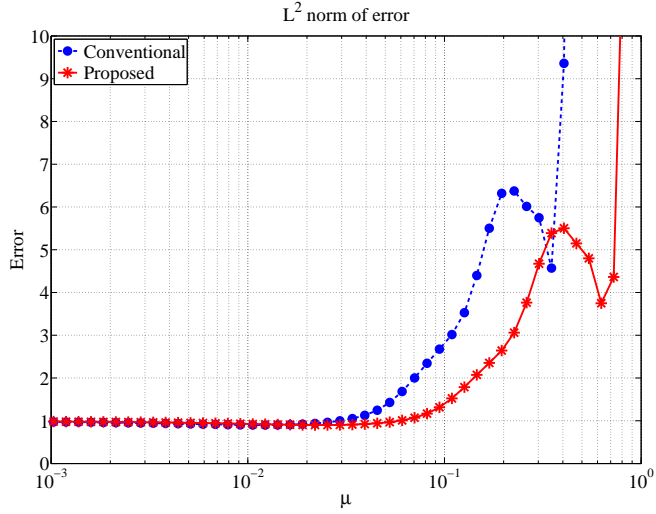}
\caption{$L^2$ norm of the error $e(t)$:
conventional (dash) and proposed (solid).}
\label{fig:mu}
\end{figure}

Fig.~\ref{fig:mu} shows the $L^2$ norm of the error $e(t)$,
$0\leq t\leq 100$ (sec) with some values of the step-size
parameter $\mu$.
Fig.~\ref{fig:mu} shows that 
the error by the proposed method is equal to 
or smaller than that by the conventional method
for almost all values of $\mu$.
Moreover,
the error by the proposed method can be small
for much wider interval than that by the conventional method.
In fact, 
the $L^2$ norm of the error $\|e\|_2<10$ if
$\mu\in(0,0.7257)$ by the proposed method,
while $\|e\|_2<10$ if $\mu\in(0,0.4051)$
by the conventional method.
That is, the interval by the proposed method is 
about 1.8 times wider than that by the conventional method.

In summary, the simulation results show that
the proposed method gives better performance for
wider interval of the step-size parameter $\mu$
on which the adaptive system is stable
than the conventional method.

\section{Conclusion}
\label{sec:conc}
In this article, we have proposed a hybrid design
of filtered-$x$ adaptive algorithm
via lifting method in sampled-data control theory.
The proposed algorithm can take account of 
the continuous-time behavior of the error signal.
We have also proposed an approximation
of the algorithm, which can be easily implemented
in DSP.
Simulation results have shown the effectiveness of
the proposed method.

\section*{Acknowledgments}
This research is supported in part by the JSPS Grant-in-Aid
for Scientific Research (B) No.\ 2136020318360203,
Grant-in-Aid for Exploratory Research No.\ 22656095,
and the MEXT Grant-in-Aid for Young Scientists
(B) No.\ 22760317.

\appendix
\section{Proof of Lemma \ref{lem:S}}
\label{app:lemma1}
First, we prove $\lambda_i\geq 0$ for $i=1,2,\ldots,N$.
Let
\[
 \vc{U}(t) = \bigl[u(t),u(t-h),\ldots,u(t-Nh+h)\bigr]^\top.
\]
Then, for non-zero vector $\vc{v}\in\Real^N$,
we have
\[
  \vc{v}^\top \Phi \vc{v} 
   = \vc{v}^\top \biggl( \int_0^\infty \vc{U}(t)\vc{U}(t)^\top \dd t\biggr)\vc{v}
   = \int_0^\infty \left|\vc{v}^\top \vc{U}(t)\right|^2 \dd t
   \geq 0.
\]
Thus $\Phi\geq 0$ and hence $\lambda_i\geq 0$ for $i=1,2,\ldots,N$.
Next, since $u(t)=0$ for $t<0$, we have
\[
  \Phi_{kl} 
   = \int_0^\infty u(t-kh)u(t-lh)\dd t
   = \int_0^\infty u\bigl(t-(k-l)h\bigr)u(t)\dd t.
\]
By Parseval's identity,
\[
 \begin{split}
  \Phi_{kl} 
   &= \frac{1}{2\pi}\int_{-\infty}^\infty 
     \overline{\hat{u}(\jj\omega)}\hat{u}(\jj\omega)\e^{\jj\omega(k-l)h}\dd\omega\\
   &= \frac{1}{2\pi}\sum_{n=-\infty}^\infty \int_{-h/\pi}^{h/\pi} 
     \left|\hat{u}\left(\jj\omega+\frac{2n\pi\jj}{h}\right)\right|^2
     \e^{\jj\omega(k-l)h}\dd\omega\\
   &= \frac{h}{2\pi}\int_{-h/\pi}^{h/\pi} S(\jj\omega)\e^{\jj\omega(k-l)h}\dd \omega.
 \end{split}
\]
Then, let $\vc{v}=[v_0,v_1,\ldots,v_{N-1}]^\top$ be a nonzero vector in $\Real^N$.
Let $\hat{v}$ denote the discrete Fourier transform of $\vc{v}$, that is,
\[
 \hat{v}(\jj\omega) := \sum_{k=0}^{N-1}v_k \e^{-\jj\omega kh},\quad \omega\in (-\pi/h,\pi/h).
\]
Perseval's identity again gives
\[
 \vc{v}^\top \vc{v} = \frac{h}{2\pi}\int_{-\pi/h}^{\pi/h} \overline{\hat{v}(\jj\omega)}\hat{v}(\jj\omega)\dd\omega.
\]
Then we have
\[
 \begin{split}
  \vc{v}^\top \Phi \vc{v}
   &= \sum_{k=0}^{N-1}\sum_{l=0}^{N-1} v_kv_l\Phi_{kl}\\
   &= \sum_{k=0}^{N-1}\sum_{l=0}^{N-1} v_kv_l\cdot\frac{h}{2\pi}\int_{-h/\pi}^{h/\pi} S(\jj\omega)\e^{\jj\omega(k-l)h}\dd \omega\\
   &= \frac{h}{2\pi} \int_{-\pi/h}^{\pi/h} S(\jj\omega)\overline{\hat{v}(\jj\omega)} \hat{v}(\jj\omega)\dd\omega\\
   &\leq \|S\|_\infty\cdot \vc{v}^\top\vc{v}.
 \end{split}
\]
It follows that
\[
 \begin{split}
  \max_{1\leq i\leq N}\lambda_i 
  &= \max\{\vc{v}^\top\Phi\vc{v}\mid \vc{v}\in\Real^N,\quad \vc{v}^\top\vc{v}=1\}\\
  &\leq \|S\|_\infty.
 \end{split}
\]

\section{Proof of Lemma \ref{lem:stability2}}
\label{app:lemma2}
Let
$\A[n] := I-\mu\Phi[n]$, $n\in\Zp$.
By the assumption 1, we have
\[
 \bigl\|\A[n]\bigr\| = \bigl\|I-\mu\Phi[n]\bigr\| \leq N+\mu\bigl\|\Phi[n]\bigr\|\leq N+\mu\gamma.
\]
Thus, the sequence $\{\A[n]\}$ is uniformly bounded.
By the assumption 2, we have
\[
 \bigl|\lambda_{\max}\bigl(\A[n]\bigr)\bigr| < 1,\quad \forall n\in\Zp.
\]
Also, by the assumption 3, we have
\[
 \bigl\|\A[n]-\A[n-1]\bigr\|\leq \epsilon,
\]
that is, the sequence $\{\A[n]\}$ is slowly varying.
With these inequalities,
the uniform exponential stability of the system \eqref{eq:auto} 
follows from Theorem 24.8 in~\cite{Rug}.

\end{document}